# Broadband wavelength-selective isotype heterojunction n$^+$-ZnO/n-Si photodetector with variable polarity

*Georgios Chatzigiannakis*[1,2], *Angelina Jaros*[3], *Renaud Leturcq*[4], *Jörgen Jungclaus*[3], *Tobias Voss*[3], *Spiros Gardelis*[2,*], *Maria Kandyla*[1,*]

[1]Theoretical and Physical Chemistry Institute, National Hellenic Research Foundation, 48 Vassileos Constantinou Avenue, 11635 Athens, Greece

[2]Department of Physics, National and Kapodistrian University of Athens, Panepistimiopolis Zografos, 15784 Athens, Greece

[3]Institute of Semiconductor Technology, Braunschweig University of Technology, Hans-Sommer Strasse 66, 38106 Braunschweig, Germany

[4]Materials Research and Technology Department, Luxembourg Institute of Science and Technology, 41 Rue du Brill, L-4422 Belvaux, Luxembourg

**Abstract**

An isotype heterojunction $n^+$-ZnO/n-Si photodetector is developed, demonstrating wavelength-selective or broadband operation, depending on the applied bias voltage. Additionally, at self-powered (zero bias) operation, it distinguishes between UV, visible, and near IR (NIR) photons by polarity control of the photocurrent. The photodetector is developed by atomic layer deposition (ALD) of ZnO on n-Si, followed by electric contact deposition and annealing. Photoluminescence measurements reveal high optical quality and improved crystallinity of annealed ZnO on silicon. Photocurrent measurements as a function of illumination wavelength and bias voltage show small negative values in the UV-visible spectral range at zero and positive bias voltage and high positive values in the NIR spectral range. For these measurements, we consider the electric contact to ZnO as the anode and the electric contact to silicon as the cathode. At negative bias voltage, the device shows broadband operation with high photocurrent values across the UV-vis-NIR.

**1. Introduction**

Zinc oxide (ZnO) is a very promising material for optoelectronic and transparent electronic applications due to its wide bandgap (3.37 eV), which is direct and therefore suitable for optoelectronics, its large exciton binding energy at room temperature (60 meV), and its high transparency (>80%) in the visible spectral range.[1]–[5] Recently, there has been a vigorous interest for the development of ZnO-based UV photodetectors which can work without bias from an external power supply.[5],[6],[7],[8] Self-powered photodetection is desirable for the development of Visible Light Communication (VLC) systems, Internet of Things (IoT) systems, triboelectric nanogenerators, *etc*.[5],[6],[9],[10] However, the development of blue-UV optoelectronic devices based on ZnO homojunctions is hindered by the difficulty in introducing reproducibly high-quality p-type impurities in ZnO, which exhibits intrinsically n-type conductivity due to bulk defects, such as oxygen vacancies and zinc interstitials.[1],[3],[4] Therefore, ZnO is usually combined with p-doped semiconductors, most commonly with silicon, which is a narrow-bandgap material and extends the operation of ZnO optoelectronic devices to the visible-near infrared (vis-NIR) spectral range. Indeed, ZnO/p-Si heterojunctions have been employed repeatedly as solar cells,[11],[12] UV-visible photodetectors,[8],[13]–[22] and LEDs,[23],[24],[25] and recently we demonstrated NIR-extended photoresponse in a laser-microstructured ZnO/p-Si photodetector.[26]

A wide variety of ZnO/p-Si photodetectors have been developed with high response in the UV-visible spectral range.[8],[13]-[22],[26] In some cases, selective detection of UV or visible light has been achieved, which is useful for filterless optoelectronic applications, such as imaging, spectroscopy, and machine vision.[27],[28],[29] Indeed, n-ZnO/p-Si photodetectors have shown either UV-blind visible detection or visible-blind UV detection, depending on the applied bias voltage.[16],[19],[21],[22] In addition to n-ZnO/p-Si heterojunctions, which operate as p-n junction minority-carrier devices, n-ZnO/n-Si isotype heterojunctions form an interesting class of majority-carrier devices. In such devices, where no minority-carrier injection is involved, a weak built-in potential barrier is formed between the two materials, which enables a two-way current flow.[30] The two-way current flow of the isotype junction is useful for wavelength-selective photodetection, as the net current changes polarity and intensity depending on the illumination wavelength, as will be shown below. This is in contrast to the more common p-n junctions, where a strong built-in potential barrier is established between the p- and n-type materials, and which operate based on an entirely different mechanism. The differences between wavelength-selective operation of isotype and p-n heterojunction photodetectors are discussed in more detail in Results and Discussion. Recently, an isotype p-SnS/p-Si device with the above mentioned characteristics has shown self-powered wavelength-selective detection.[31] However, n-ZnO/n-Si heterojunctions have been studied less extensively regarding their optoelectronic properties and their potential for detection [6],[32]–[35] and, to the best of our knowledge, wavelength-selective photodetection with these devices has not been demonstrated until now.

In this work, we develop an isotype heterojunction $n^+$-ZnO/n-Si photodetector, which, depending on the bias voltage, shows for the first time either UV-visible blind NIR detection or broadband UV-vis-NIR detection without the need of additional optical elements, such as filters. Specifically, it demonstrates a high NIR responsivity at zero and positive bias voltage and a broadband responsivity at negative bias voltage. This feature is complementary to the existing n-ZnO/p-Si selective photodetectors, which detect either UV or visible radiation but do not distinguish between NIR and UV/visible radiation.[16],[19],[21],[22] Additionally, there is a photocurrent polarity switch for positive and zero bias voltages, depending on the illumination wavelength. The device exhibits small negative photocurrent values in the blind UV-vis spectral range and high positive photocurrent values in the responsive NIR spectral range, thereby distinguishing between the UV, visible, and NIR photons by polarity control. On the other hand, at negative bias there is no photocurrent polarity switch or wavelength-selective responsivity. The device has been developed by atomic layer deposition (ALD) of ZnO on silicon. Other cost-effective methods, such as the sol-gel process, have also been widely used for the growth of ZnO or other metal oxides[36],[37]. ALD provides high-quality thin films at low deposition temperatures, high conformality, large-scale uniformity, and monoatomic layer thickness control.[38] The novelty of this work is the development, for the first time, of a wavelength-selective photodetector based on the functionality of the majority-carrier device $n^+$-ZnO/n-Si that functions either in the UV-vis-NIR or NIR spectral range, depending on the applied bias voltage. Moreover, even at self-powered (zero bias) conditions, the device shows wavelength selectivity, determined by the polarity of the photocurrent. Such a device may find potential use in filterless optoelectronic applications,[27],[28],[29] even at self-powered conditions, and be easily included in silicon-based photonic integrated circuits, as it is developed on a silicon wafer.[39]

## 2. Experimental Section

<u>Development of n$^+$-ZnO/n-Si devices.</u> ZnO thin films (200 nm nominal thickness) were deposited on n-type monocrystalline (111) silicon substrates (phosphorus doped with carrier concentration $N_d$ = 2-3*10$^{15}$ cm$^{-3}$) by ALD, forming ZnO/Si heterojunctions. ZnO deposition was performed in a Beneq TFS200 reactor at a chamber temperature of 150°C, using diethylzinc (99.99%, Sigma Aldrich) and deionized water as precursors. Prior to deposition, the silicon substrates were etched in an aqueous solution of hydrofluoric acid (5% HF for 5 min) to remove the native SiO$_2$. ZnO films deposited on glass have been characterized by Hall effect measurements. As-grown ZnO films are n-type with a carrier density $n$ = (2.5 ± 0.2)*10$^{19}$ cm$^{-3}$ and a mobility $\mu_n$ = 32 ± 5 cm$^2$ (V s)$^{-1}$. Top finger contacts were deposited on ZnO by thermal evaporation of Al (220 nm) via a shadow mask, in order to allow the photoexcitation of the diode. Bottom electrodes were deposited on the back surface of silicon by thermal evaporation of Ag (250 nm). The entire device was thermally annealed (300 °C, 40 min) in inert pure nitrogen ambient (N$_2$), which is preferred compared to air, in order to reduce the Ag/Si contact resistance and improve its ohmic behavior without oxidation. Indeed, the signal of the device was significantly improved after thermal annealing. From Hall measurements of the ZnO films on glass substrates, we find that after annealing in N$_2$ atmosphere at 300°C for 40 min, both the carrier concentration and mobility decrease by one order of magnitude, with $n_{annealed}$ = (3.0 ± 0.5)*10$^{18}$ cm$^{-3}$ and $\mu_{n,annealed}$ = 2.0 ± 0.5 cm$^2$ (V s)$^{-1}$.[26] Low-temperature annealing, which is useful for large-area applications, will be the subject of a future study.

Characterization. The heterojunction morphology was investigated with a field-emission Scanning Electron Microscope (SEM). Room-temperature and low-temperature (4 K) photoluminescence (PL) measurements were carried out with a HeCd laser ($\lambda$ = 325 nm, $P$ = 1 mW). The integration time was 200 ms at 4 K and 500 ms at room temperature, except for the measurement shown in Figure 2a, where the integration time was 2 s.

Electrical and optoelectronic measurements. *I-V* measurements were carried out in dark and under illumination in order to investigate the optoelectronic behavior of the photodetector. A bias voltage was applied using a Keithley 230 voltage source and the current was monitored by a Keithley 485 picoammeter. Illumination of the heterojunction for photocurrent measurements was performed with a xenon lamp as a light source, monochromated through an Oriel ¼. 77200 monochromator in the 365 – 1000 nm spectral range. The positive terminal of the voltage source was connected to the top Al contact on ZnO, which was considered as the anode, and the negative terminal to the bottom Ag contact on silicon (inset of Figure 4a). Therefore, all voltage values presented in this work refer to the difference of the ZnO potential with respect to the silicon potential. Capacitance-voltage (*C-V*) measurements of the heterojunction were performed in dark conditions by a Boonton 72B capacitance meter at a frequency of 1 MHz. The responsivity of the photodetector was calculated by normalizing the photocurrent to the power of the xenon lamp for each wavelength, using a calibrated, enhanced-UV, commercial silicon photodiode (Thorlabs SM1PD2A).

## 3. Results and Discussion

### 3.1 Structure and Surface Morphology

A cross-sectional SEM micrograph of the device, shown in **Figure 1**, reveals that the thickness of the ZnO thin film grown by ALD on silicon is ~210 nm. The crystalline structure of ZnO thin films developed under identical conditions (same equipment, same conditions) was investigated by X-ray diffraction (XRD) analysis.[40] The results indicated textured polycrystalline films with a preferential orientation of the crystal grains with the (100) crystallographic planes parallel to the surface. The grains of similar ZnO films on silicon have been found to have an elongated shape with a typical length 30 – 50 nm.[26]

**3.2 Photoluminescence**

**Figures 2**a and **2**b show the PL spectra of ALD-grown ZnO before and after thermal annealing (300 °C, $N_2$ gas ambient, 40 min), respectively, obtained at room temperature. As a reference, the room-temperature PL spectrum of a ZnO single-crystal wafer (Crystec, *m*-plane orientation) is shown in the inset of Figure 2b. The ZnO single crystal shows a strong UV peak at 380 nm and a weaker green band centered at 528 nm. The UV emission originates from exciton and electron-hole pair recombination processes while the green band is attributed mainly to ionized or neutral oxygen vacancies and also to other surface or bulk defects like zinc interstitials, oxygen antisites, *etc*.[26],[41] The as-grown ZnO shows a UV peak at 391 nm and a broad visible green band centered at 495 nm (Figure 2a). Compared to the single-crystal ZnO, the UV peak is red-shifted and broader while the defect-related band is also significantly broader and of higher relative intensity. After thermal annealing, we observe a UV peak at 384 nm and a weaker green band centered at 523 nm (Figure 2b), similar to the spectrum of ZnO single-crystal wafer, as shown in the inset of Figure 2b. Compared to the as-grown sample, the defect-related band is weaker and narrower, indicating enhanced optical quality of ZnO since thermal annealing results in the deactivation of optical-active defects.[26] The reduction of defects after annealing is compatible with the observed reduction of carrier concentration (see Section 2), since free carriers are known to originate from zinc interstitials and oxygen vacancies in ZnO. Regarding the UV peak, it becomes narrower and blue-shifted upon annealing, resembling more the single-crystal UV peak, due to improved crystalline quality. Based on the UV peak of the annealed ZnO sample, and taking into account the ZnO exciton binding energy at room temperature (60 meV),[26],[41] we estimate an optical bandgap of 3.3 eV.

Figure 2c shows the PL spectrum of the annealed ALD ZnO sample at low-temperature conditions (4 K), where we observe a strong UV peak at 374 nm and a green band at 475 nm. The UV peak is blue-shifted, narrower, and more intense, while the defect-related band is also blue-shifted, compared to room-temperature PL spectra. The intensity increase and narrowing of the UV peak as well as the blue shift, which is due to an increased bandgap at low-temperature conditions, are expected, as reported in literature.[41] In the inset of Figure 2c, we can resolve three different sub-peaks in the main UV peak at 4 K, which can be attributed to the recombination of excitons bound to donors, longitudinal optical (LO) phonon replicas, two-electron transitions, *etc*.[26],[41]

### 3.3 Electrical and Optoelectronic Operation

In order to verify that the Al and Ag electrodes form ohmic contacts with ZnO and silicon, respectively, we show in **Figure 3**a current-voltage (*I-V)* measurements between two Al electrodes deposited on ZnO as well as between two Ag electrodes deposited on silicon. We observe that the *I-V* curves are linear, indicating an ohmic behavior. The ohmic Al-ZnO contact is attributed to the formation of an extremely small Schottky barrier, which enables carriers to be transferred to both sides, since the electron affinity of ZnO is comparable to the work function of Al.[42],[43] In the case of Ag-Si, the ohmic behavior of the contact is achieved by the thermal annealing process, described in Section 2.

**Figure 3**b shows $1/C^2$ *vs*. voltage (Mott-Schottky plot) for the n$^-$-ZnO/n-Si heterojunction. We observe that $1/C^2$ decreases with a decreasing reverse bias voltage until it reaches a minimum quasi-steady value under forward bias. The linear region in the Mott-Schottky plot indicates an abrupt heterojunction between silicon and ZnO. The extraction of the linear region

to the voltage axis (where $1/C^2 = 0$) indicates the built-in potential of the heterojunction, $V_{bi} = 0.57$ V, which is comparable to literature $V_{bi}$ values for ZnO/n-Si heterojunctions.[34,44]

The optoelectronic behavior of the device was investigated by *I-V* measurements in dark and under illumination at different spectral ranges. **Figure 4**a shows *I-V* characteristics under illumination with visible light (500 nm wavelength) and UV light (380 nm wavelength), compared to those carried out in the dark. We observe that the *I-V* curve in dark shows a rectifying behavior with significantly higher current values in the negative voltage region (negative potential on n+-ZnO with respect to n-Si) than in the positive voltage region, with a rectification ratio $I_-/I_+ \sim 100$ at -1 V. Under 380 nm and 500 nm illumination, a photocurrent is observed in the negative voltage region. **Figure 4**b shows the *I-V* characteristics in dark and under NIR illumination (1000 nm wavelength). Interestingly, in this case a photocurrent is observed in the positive voltage region in contrast to 380-nm and 500-nm illumination conditions. **Figure 4**c shows the semilog counterparts of Figures 4a and b. Under 500-nm illumination, we clearly observe a shift of the *I-V* curve towards the positive voltage region as a result of the photovoltaic effect,[45] indicating a positive value of open-circuit voltage ($V_{oc}$). Under 380-nm illumination there is also a small positive shift of the *I-V* curve, as shown in the inset of **Figure 4**c, indicating a smaller $V_{oc}$ value compared to 500-nm illumination. However, under 1000-nm illumination we observe a shift towards the negative voltage region, indicating a negative value of $V_{oc}$.

**3.4 Spectral Responsivity**

Photocurrent measurements elaborate on the spectral response of the photodetector. **Figure 5**a shows the spectral responsivity of the n+-ZnO/n-Si device at short-circuit conditions (0 V). The algebraic values of the responsivity are plotted, therefore a negative responsivity corresponds to a negative photocurrent in the circuit, according to the convention used in the insets of **Figures 4**a and b, *i.e.*, a negative photocurrent flows from silicon to ZnO in the heterojunction. On the other hand, a positive responsivity corresponds to a positive photocurrent, flowing from ZnO to silicon in the heterojunction. The responsivity is near zero in the UV-vis spectral range and increases sharply in the NIR. A magnification of the UV-vis responsivity, shown in the inset of **Figure 5**a, reveals negative responsivity values for 365 – 638 nm, which correspond to a negative short-circuit photocurrent ($I_{sc}<0$), and increasingly positive responsivity values for $\lambda > 638$ nm, which indicate an inversion of the short-circuit photocurrent direction ($I_{sc}>0$). The negative photocurrent effect was investigated for different incoming light intensities, revealing increasingly negative $I_{sc}$ values in the spectral range 365-638 nm for increasing light intensities (Supporting Information, **Figure S**1). Therefore, at zero bias the device demonstrates selective UV-vis blind NIR detection accompanied by a variable photocurrent polarity, depending on the illumination wavelength. Similarly, wavelength-selective photodetection based on flipping photocurrent polarity has also been reported for an isotype (p-p) SnS/Si photodetector, operating at zero bias with high speed. [31]

**Figure 5**b shows the spectral responsivity at +1 V bias on ZnO with respect to silicon. Similar to short-circuit conditions, we observe a near-zero responsivity for 365 – 498 nm, where small negative responsivity values are recorded (inset of **Figure 5**b), while for $\lambda > 498$ nm we observe increasingly positive responsivity values, indicating an inversion of the photocurrent direction. We note that the negative responsivity spectral range is narrower at +1 V bias, compared to 0 V bias (inset of **Figure 5**a). Therefore, under positive voltage bias the device demonstrates selective UV-vis blind NIR detection accompanied by variable polarity of the photocurrent, depending on the illumination wavelength, similar to the 0 V case.

**Figure 5**c shows the spectral responsivity at -1 V bias applied on ZnO with respect to silicon. In contrast to the previous two cases, we observe high responsivity values across the entire UV-vis-NIR spectral range. More specifically, we observe three distinct spectral regions

in the responsivity spectrum. Region A, centered at 388 nm, is attributed to band-to-band transitions in ZnO. Region B, centered at 527 nm, is attributed to transitions from energy states within the bandgap of ZnO, due to bulk defects, but also to band-to-band transitions in silicon. Region C, centered at 910 nm, is attributed to band-edge absorption in silicon. We note that the centers of peaks A and B correlate well with the PL emission peaks of annealed ZnO at room temperature, due to band-to-band and defect transitions, respectively (**Figure 2**b). The algebraic values of the responsivity are negative across the entire spectral range shown in **Figure 5**c, however for simplicity we show the absolute values. Therefore, at -1 V the device demonstrates broadband detection in the UV-vis-NIR spectral range, in contrast to zero or positive voltage bias conditions (Supporting Information, **Figure S**2). Interestingly, the responsivity in this case is spectrally similar to that observed in our previous work for ZnO/p-Si photodetectors.[26] In addition, we evaluated the specific detectivity ($D^*$) of the isotype device at 0 V [46],[47], -1 V, +1 V [48]. The detectivity values, which correspond to the maximum value of responsivity, are of the order of $10^{11}$-$10^{12}$ Jones (Supporting Information, **Table S**1), in agreement with other studies on ZnO/Si photodetectors.[48]

### 3.5 Operation Mechanism

In order to understand the operation of the $n^+$-ZnO/n-Si photodetector, we start by the formation of the heterojunction. **Figure 6** shows the energy band diagram of the $n^+$-ZnO/n-Si heterojunction in dark conditions. The Fermi level of ZnO lies within the conduction band[49] because its carrier density ($3*10^{18}$ cm$^{-3}$) is higher than the calculated effective density of states in the ZnO conduction band ($N_c = 2*10^{18}$ cm$^{-3}$), assuming an effective electron mass of $0.19*m_o$, where $m_o$ is the free-electron mass.[46] When the junction is formed, and until the establishment of thermal equilibrium, electrons diffuse from ZnO towards silicon, because the carrier concentration in $n^+$-ZnO is three orders of magnitude higher than in n-Si. Thus, a small depletion region (positive space charge) is formed in ZnO and an accumulation region (negative space charge) in silicon, as shown in **Figure 7**. The formation of this n$^+$-n junction results in the *I-V* curves in dark conditions shown in **Figure 4**, which demonstrate the diode conducts when a negative potential is applied to n$^+$-ZnO with respect to n-Si.

For the operation of the photodetector under illumination, we examine first the zero-bias case ($V = 0$). **Figure 7**a is a schematic of the n$^+$-ZnO/n-Si heterojunction, which shows the ZnO depletion region and the silicon accumulation region at short-circuit conditions. Under UV illumination on the ZnO side, photoelectrons are generated mostly in the conduction band of ZnO. Some of these photogenerated electrons cross over the built-in potential barrier towards silicon, resulting in the generation of a small negative photocurrent, according to the electric circuit convention we follow in this study (**Figure 4**a inset). A similar behavior has been observed for photogenerated holes in an isotype heterojunction p-SnS/p-Si photodetector.[31] This is manifested in the UV part of the short-circuit responsivity, shown in the inset of **Figure 5**a, which is negative and decreases monotonically with wavelength until it reaches a negative maximum at dip A = 388 nm. This dip correlates well with the UV peak in the ZnO PL spectrum (384 nm, **Figure 2**b) and Region A of the responsivity at -1 V (388 nm, **Figure 5**c). For $\lambda >$ 388 nm, electrons are photogenerated in both materials and there is a fine interplay between two small opposite currents, originating from ZnO and silicon (**Figure 7**a, top). Visible photons generate photoelectrons in ZnO due to transitions from defect states within the energy bandgap, which result in a negative photocurrent as they cross the potential barrier towards silicon. At the same time, visible light that reaches silicon promotes electrons in the conduction band due to band-to-band transitions. These photogenerated electrons drift towards ZnO due to the built-in electric field of the heterojunction, resulting in the generation of a positive photocurrent, according to our convention (**Figure 4**a inset). The short-circuit responsivity in the inset of **Figure 5**a shows that in the spectral range 365 – 638 nm the net photocurrent (and hence the responsivity) remains negative, indicating that the photoelectrons generated in ZnO dominate those generated in silicon. We note that the turning point for the negative responsivity is at dip B = 504 nm, which correlates well with the defect related-band in the ZnO PL spectrum (523 nm, **Figure 2**b). Beyond this wavelength, visible-light absorption starts decreasing in ZnO and the negative component of the photocurrent, originating from ZnO photoelectrons, becomes less and less dominant, until the net photocurrent (and responsivity) becomes positive for $\lambda >$ 638 nm, where silicon photoelectrons dominate. Under NIR illumination, ZnO acts as a transparent window and photoelectrons are generated mainly in silicon (**Figure 7**a, bottom). These electrons are transported towards ZnO by the built-in electric field of the heterojunction, resulting in a high and positive photocurrent, and a sharply increasing responsivity, for $\lambda >$ 800 nm (**Figure 5**a). **Figure 7**b summarizes the inversion of photocurrent polarity, depending on the illumination wavelength, for zero bias. For 365 nm $< \lambda <$ 638 nm, the small net photocurrent $I_{ph,UV-vis}$, represented by the rainbow-colored arrow, is negative, indicating the ZnO-generated component, $I_{ph,ZnO}$, prevails. For $\lambda >$ 638 nm, the net photocurrent $I_{ph,vis-NIR}$, represented by the red arrow, is positive, indicating the silicon-generated component, $I_{ph,Si}$, prevails and results in high photocurrent and responsivity.

The photocurrent polarity switch also results in the polarity switch of $V_{oc}$, shown in **Figure 4**c. The negative photocurrent generated in the spectral range 365 – 638 nm becomes zero when a positive bias voltage is applied to ZnO with respect to silicon. For this reason, we observe positive $V_{oc}$ values under 380 nm and 500 nm illumination wavelengths. On the other hand, the positive photocurrent generated in the spectral range 638 – 1000 nm becomes zero when a negative bias voltage is applied to ZnO with respect to silicon. Indeed, we observe a negative $V_{oc}$ under 1000 nm illumination wavelength.

The operation of the n$^+$-ZnO/n-Si photodetector at positive bias voltage (positive voltage on ZnO with respect to silicon) is depicted in **Figure 7**c. The situation is similar to the zero bias case, with an extended depletion/accumulation region because the external electric field is superimposed on the built-in field of the heterojunction. Under UV-visible illumination, the photogenerated electrons in ZnO have to cross over an increased potential barrier towards silicon. At the same time, photogenerated electrons in silicon drift towards ZnO by an increased electric field. As a result, at +1 V the net photocurrent ($I_{ph,UV\text{-}vis} = I_{ph,Si} - I_{ph,ZnO}$), and corresponding responsivity, remain negative for a decreased spectral range (365 – 498 nm), compared to zero bias conditions (inset of **Figure 5**b). Interestingly, the wavelength for which the photocurrent becomes positive (498 nm) is similar to the wavelength of the turning point of the responsivity at $V = 0$ (504 nm, inset of **Figure 5**a) and they both correlate well with the defect related-band in the ZnO PL spectrum (523 nm, **Figure 2**b). Under NIR illumination, ZnO acts again as a transparent window and photogenerated electrons in silicon are transported towards ZnO by an increased electric field, resulting in a positive photocurrent and a sharply increasing responsivity for $\lambda > 800$ nm, which is higher than the zero-bias NIR responsivity (**Figures 5**b and S2). When a higher positive bias voltage is applied to ZnO with respect to silicon, the positive component of the photocurrent, from silicon towards ZnO, dominates even for wavelengths in the UV-visible spectral range, due to an increased potential barrier, which almost eliminates the diffusion current from ZnO towards silicon. Thus, at +2 V the responsivity takes positive values across the entire UV-vis-NIR spectral range (Supporting Information, **Figure S**3).

The situation is entirely different when a negative bias voltage is applied to ZnO with respect to silicon (**Figure 7**d). Now the external electric field is opposite to the built-in electric field of the heterojunction and dominates. The depletion/accumulation regions of the heterojunction shrink. Therefore, photogenerated electrons either at ZnO or silicon are transported by the external electric field in the same direction, which corresponds to a negative photocurrent in our convention. Photocurrents from ZnO and silicon are no longer opposite but they flow in the same direction. Indeed, for this case we measure an entirely negative responsivity, the absolute value of which we show in **Figure 5**c, for simplicity. The algebraic values of responsivity for selected wavelengths are shown in **Figure S**4, Supporting Information. There we can see that NIR responsivity at 1000 nm switches polarity following the polarity of the bias voltage, while visible and UV responsivities are almost zero for positive bias voltages. Therefore, at -1 V, the photodetector shows a high and broadband responsivity (**Figure 5**c), similar to the n-ZnO/p-Si case, which has a strong p-n junction built-in electric field.[26]

We note that the fine opposite photocurrent interplay, which leads to the particular wavelength selectivity in our device, would not be able to occur in a non-isotype p-n heterojunction with a strong built-in electric field, which restricts current flow to only one direction. Wavelength-selective n-ZnO/p-Si photodetectors usually show UV response in forward bias, due to UV-generated photoelectrons in ZnO drifting under the influence of the bias electric field to silicon, and suppressed visible response because vis-photogenerated holes in p-Si cannot drift to ZnO due to the big valence band offset between the two materials.[16],[19],[21] An isotype heterojunction device, as the one described in this work, does not rely on drift behaviors of both electrons and holes and, furthermore, offers a weak built-in potential barrier, which can sustain opposite currents and a wavelength selectivity in a different spectral range than n-ZnO/p-Si devices.

## 4. Conclusions

In conclusion, we developed for the first time a wavelength-selective n+-ZnO/n-Si photodetector, which detects either the entire UV-vis-NIR or only the NIR spectral range, depending on the applied bias voltage. Specifically, when positively or zero biased, it can selectively detect NIR irradiation, whereas when negatively biased it operates as a broadband UV-vis-NIR photodetector. Moreover, at self-powered conditions (zero bias) and also at positive voltage bias, the photodetector demonstrates polarity control of the photocurrent, *i.e.*, negative photocurrent for UV-vis and positive photocurrent for NIR illumination. The observed wavelength selectivity is enabled by the isotype nature of the device, which creates a weak built-in potential barrier that can sustain a fine interplay between two opposite photocurrents and the fact that illumination within a range of wavelengths can excite selectively ZnO or silicon or both materials of the heterojunction. These results pave the way for the fabrication of low-cost isotype photodetectors, which, as majority-type devices, are expected to show high response speed, with both broadband operation and wavelength selectivity for filterless, silicon-based optoelectronic applications.

**Acknowledgments**

We acknowledge support of this work by the project "Advanced Materials and Devices" (MIS 5002409) which is implemented under the "Action for the Strategic Development on the Research and Technological Sector", funded by the Operational Programme "Competitiveness, Entrepreneurship and Innovation" (NSRF 2014-2020) and co-financed by Greece and the European Union (European Regional Development Fund). We gratefully acknowledge support by the Deutsche Forschungsgemeinschaft (DFG, German Research Foundation) Research Training Group GrK1952/2 'Metrology for Complex Nanosystems'.

**Figure captions**

**Figure 1:** Cross-sectional SEM image of the $n^+$-ZnO/n-Si heterojunction. The thickness of the ZnO thin film on silicon is ~210 nm.

**Figure 2:** Room-temperature photoluminescence spectra of ZnO on n-Si (a) before and (b) after thermal annealing (300 °C, $N_2$ gas ambient, 40 min). Inset shows the room-temperature

photoluminescence spectrum of a ZnO single-crystal wafer. (c) Low-temperature photoluminescence spectrum of annealed ZnO on n-Si. Inset shows the UV peak in high resolution.

**Figure 3:** (a) *I-V* characteristics of Al electrodes on ZnO and Ag electrodes on Si. (b) $1/C^2$ *vs.* voltage plot of the $n^+$-ZnO/n-Si device (Mott-Schottky plot). The extraction of the linear region to the voltage axis (where $1/C^2 = 0$) indicates the built-in potential of the heterojunction, $V_{bi} = 0.57$ V.

**Figure 4:** *I-V* characteristics of the $n^+$-ZnO/n-Si photodetector in dark conditions and under (a) UV and visible illumination and (b) NIR illumination. (c) Semilog plot of (a) and (b). The inset in (c) shows the semilog plot under UV illumination, magnified.

**Figure 5:** Responsivity of the $n^+$-ZnO/n-Si photodetector in the UV-vis-NIR spectral range at (a) short-circuit conditions (0 V), (b) +1 V bias on ZnO with respect to silicon, and (c) -1 V bias on ZnO with respect to silicon (absolute value).

**Figure 6:** Energy band diagram of the $n^+$-ZnO/n-Si heterojunction in equilibrium. $E_c$ is the minimum of the conduction band, $E_v$ is the maximum of the valence band, $E_F$ is the Fermi level, $E_g$ is the energy bandgap, and $\chi$ is the electron affinity of the materials.

**Figure 7:** (a) Schematics showing the $n^+$-ZnO/n-Si heterojunction at zero bias under UV (top), visible (top), and NIR (bottom) illumination. Black circles: electrons, yellow circles: photogenerated electrons. Electron depletion and accumulation regions in ZnO and silicon, respectively, are also shown. (b) The direction of net photocurrent at zero bias under UV-visible illumination ($I_{ph,UV\text{-}vis}$, rainbow-colored arrow) and visible-NIR illumination ($I_{ph,vis\text{-}NIR}$, red arrow). $I_{ph,ZnO}$ and $I_{ph,Si}$ are photocurrent components generated in ZnO and silicon, respectively. (c)

Photocurrent direction for positive bias voltage on ZnO with respect to silicon and (d) for negative bias voltage on ZnO with respect to silicon, for different illumination wavelengths.

Figure 1

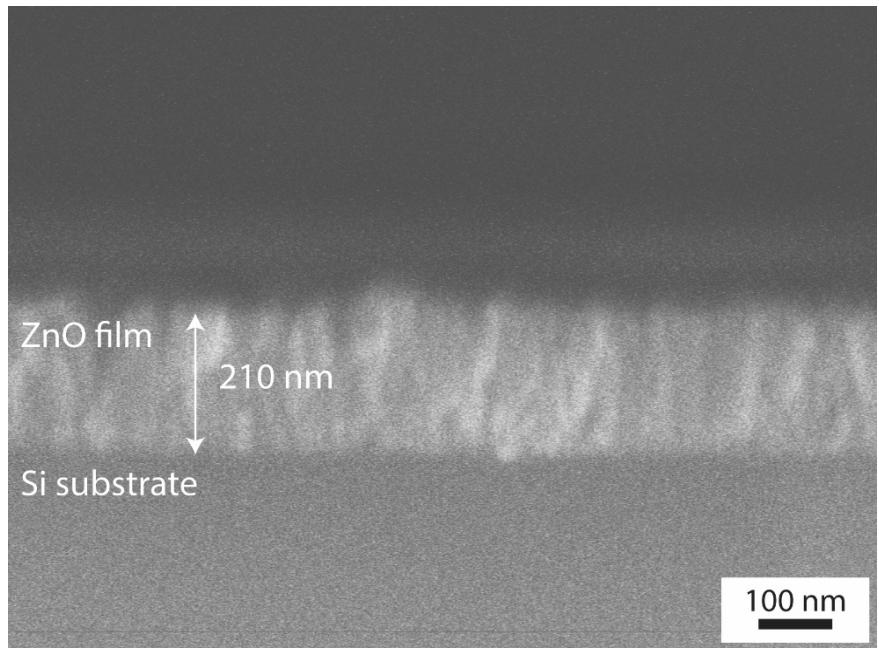

Figure 2

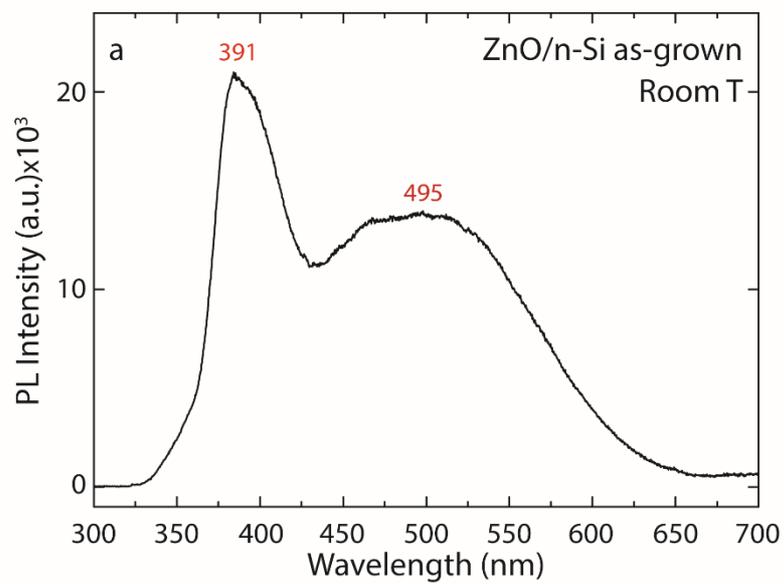

Figure 2a

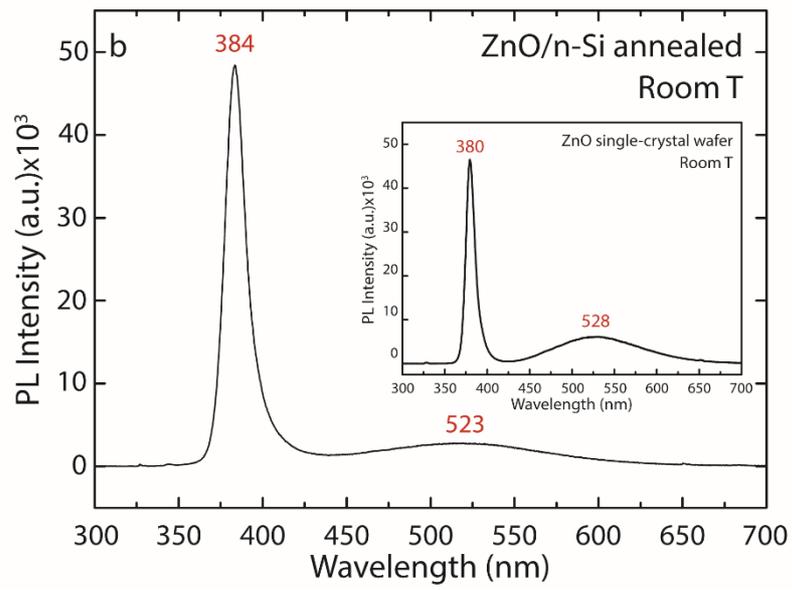

Figure 2b

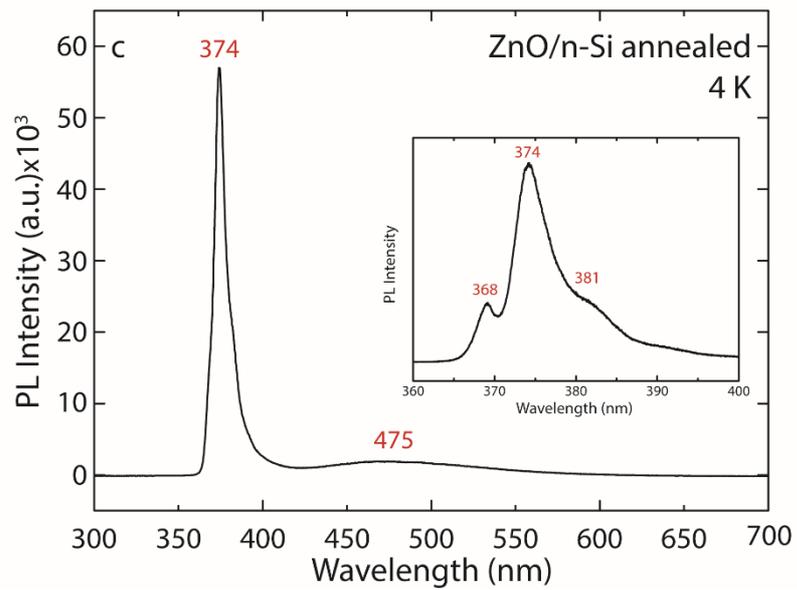

Figure 2c

Figure 3

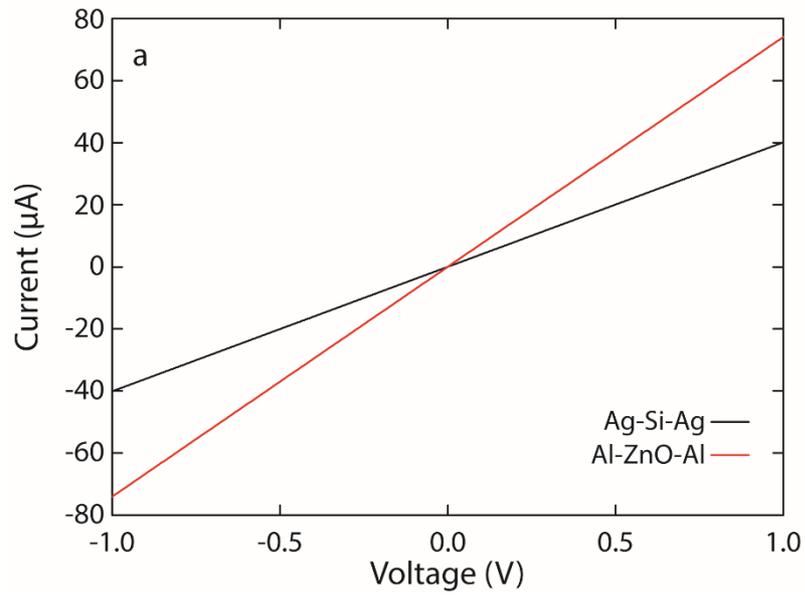

Figure 3a

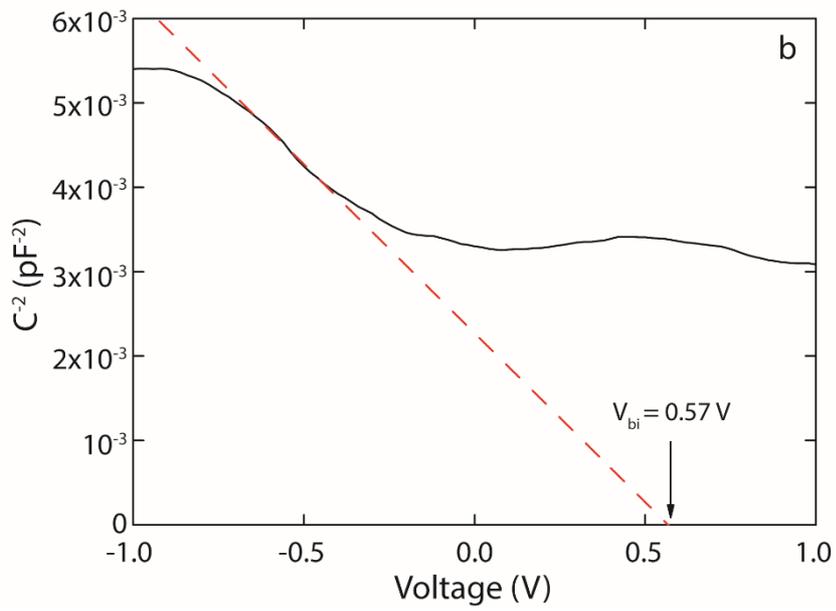

Figure 3b

Figure 4

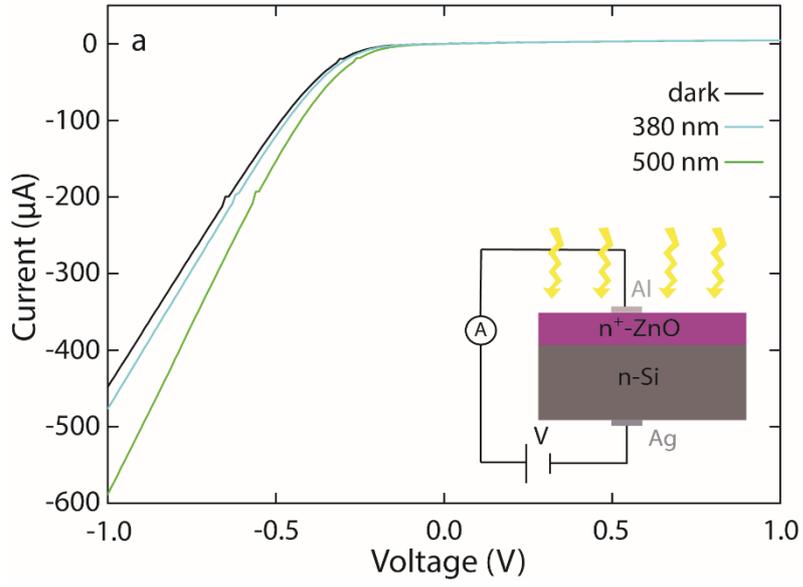

Figure 4a

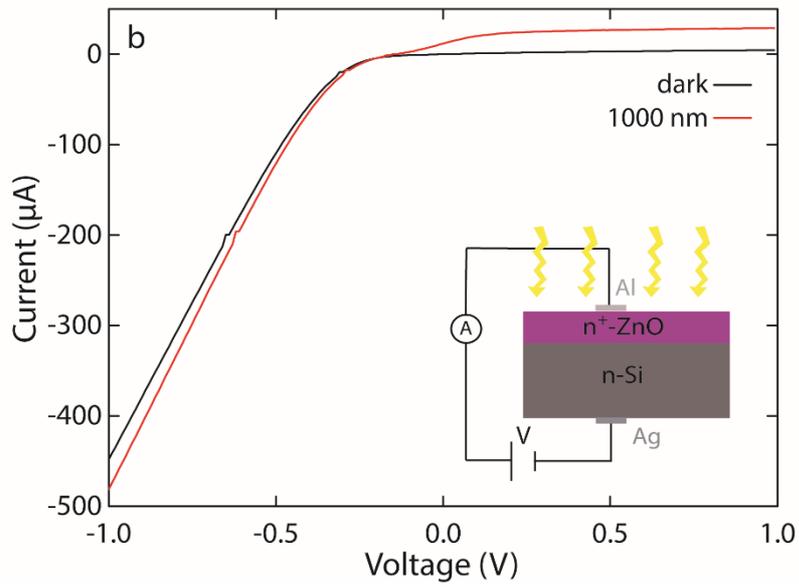

Figure 4b

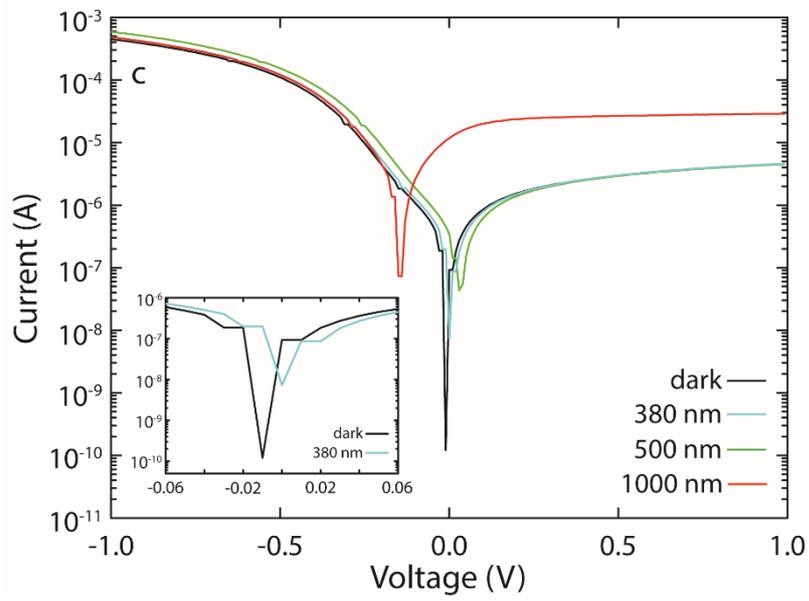

Figure 4c

Figure 5

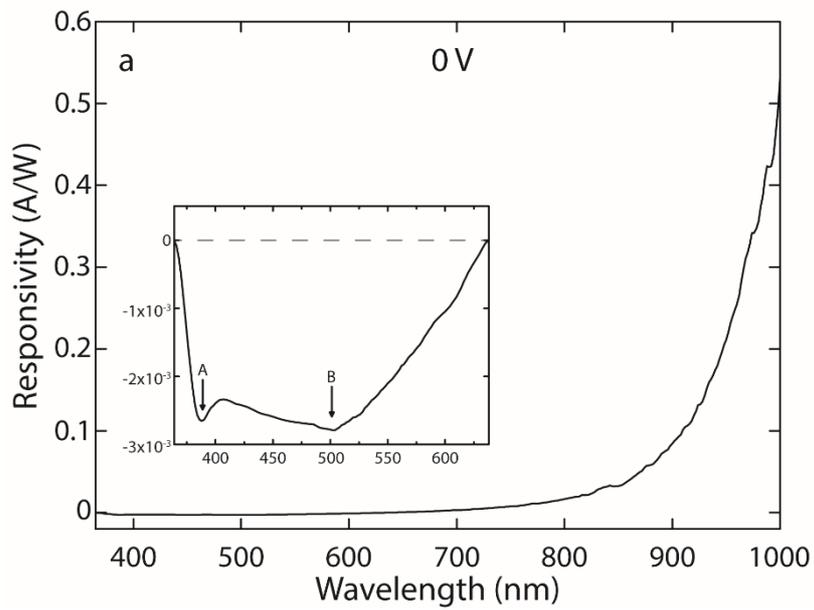

Figure 5a

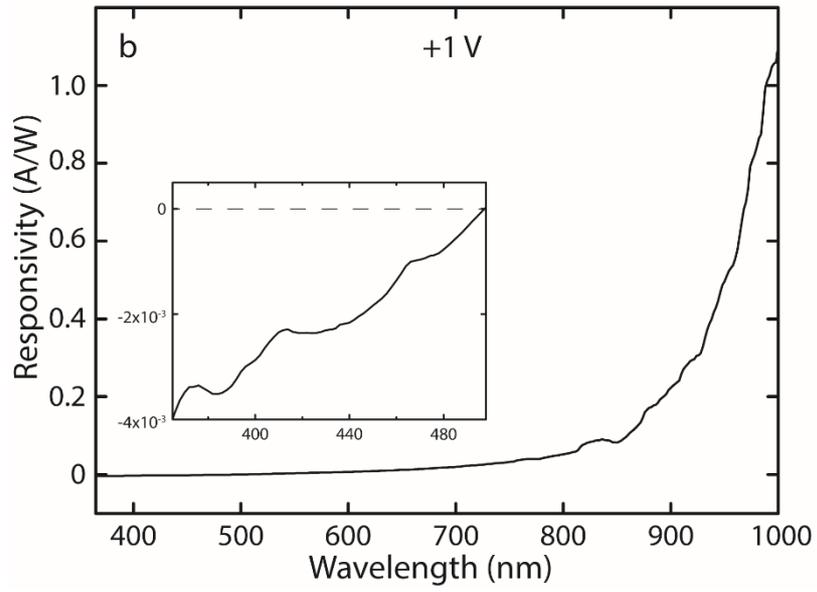

Figure 5b

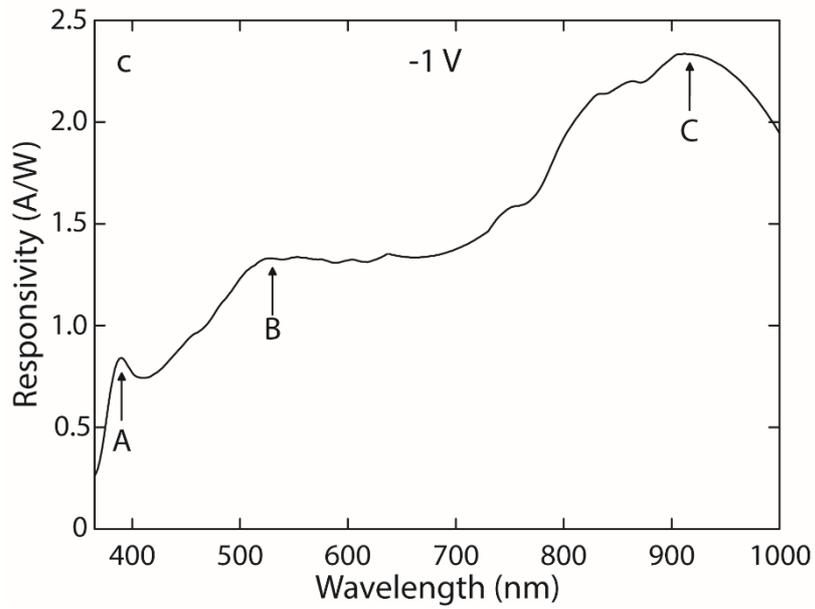

Figure 5c

Figure 6

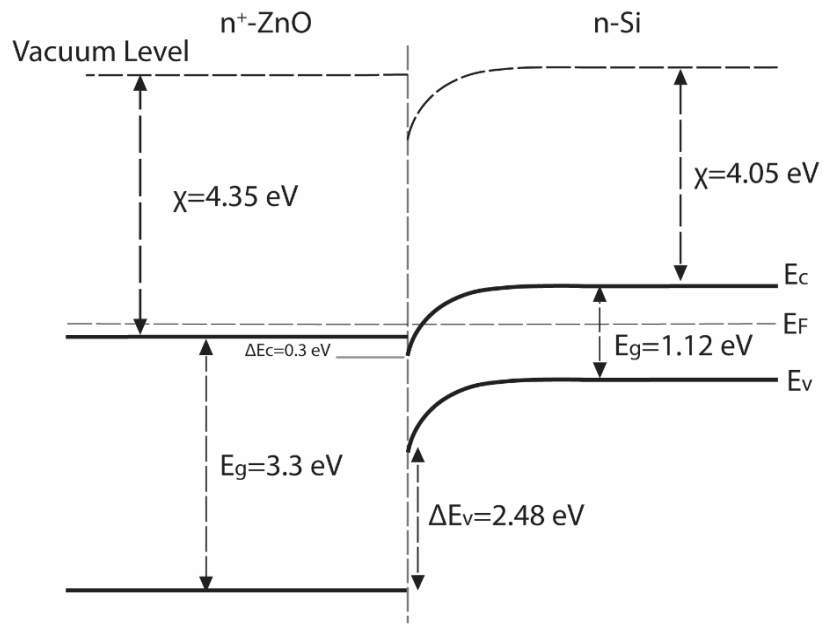

Figure 7

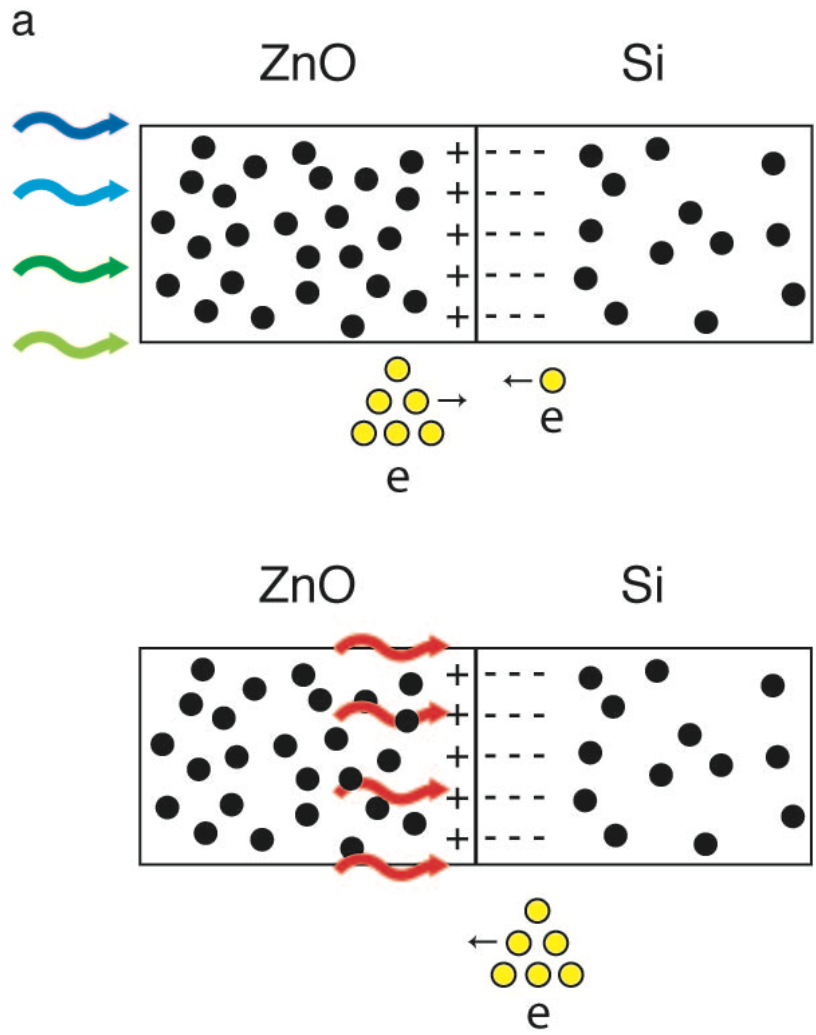

Figure 7a

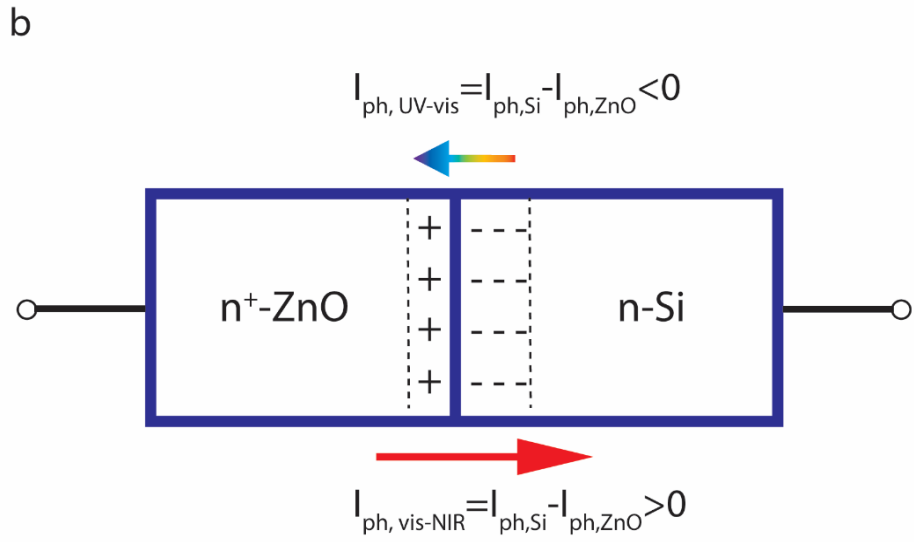

Figure 7b

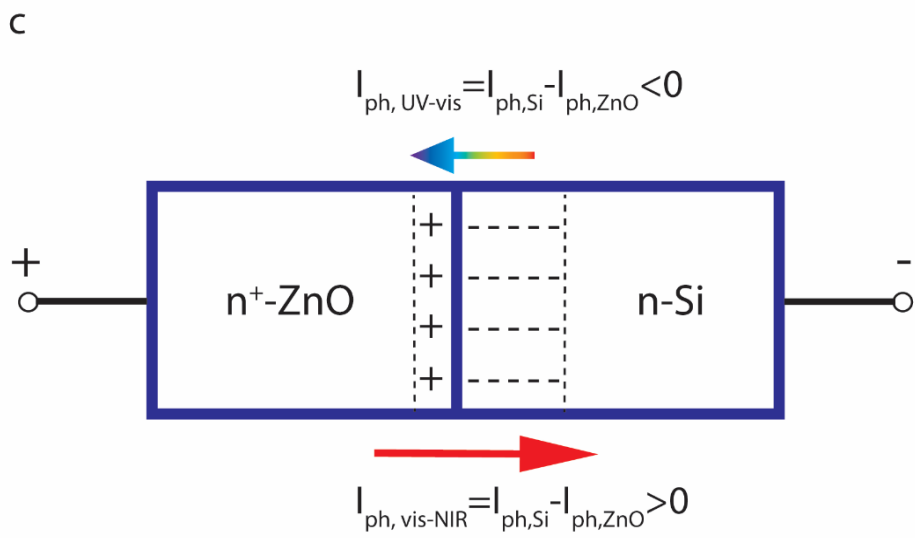

Figure 7c

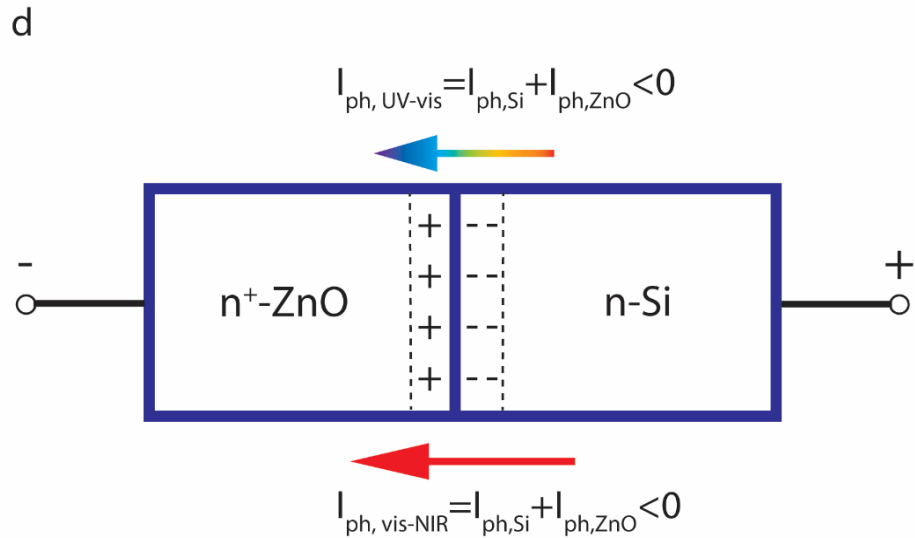

Figure 7d